\documentclass{blois}

\usepackage{float}

\def\Journal#1#2#3#4{{#1} {\bf #2}, #3 (#4)}

\def\NIMA{{\em Nucl. Instrum. Methods} A}

\def\PRL{\em Phys. Rev. Lett.}
\def\PRD{{\em Phys. Rev.} D}

\def\APJ{{\em ApJ.}}

\def\be{\begin{equation}}
\def\ee{\end{equation}}
\def\bea{\begin{eqnarray}}
\def\eea{\end{eqnarray}}

\newcommand{\Photo}{}

\begin{document}
\vspace*{4cm}
\title{Diffuse Supernova Neutrino Background}

\author{ A.D. Santos }

\address{\'Ecole Polytechnique, IN2P3-CNRS, Laboratoire Leprince-Ringuet, \\ F-91120 Palaiseau, France}

\maketitle\abstracts{
The Diffuse Supernova Neutrino Background (DSNB) is the collection of neutrinos from all core-collapse supernovae (CCSNe) since the beginning of the universe. It is sensitive to the universe's stellar formation history, the fraction of CCSNe forming black holes, and cosmological expansion. To this date, it has yet to be detected. The most sensitive experimental search is from the Super-Kamiokande experiment, and the next few years will see other sensitive experiments like Hyper-Kamiokande and the Jiangmen Underground Neutrino Observatory come online. Here, we summarize the latest results and sensitivity for the DSNB search as well as its potential to probe new physics.}

\section{Introduction}

About once every second in the observable universe, a star undergoes the cataclysmic explosion of a core-collapse supernova (CCSN). These CCSNe are the fate of massive stars ($\geq 8M_\odot$) once fusion in their cores reaches iron. No further radiation-producing fusion is possible at this stage, and collapse begins once enough iron has been produced such that neither the radiation pressure nor the electron degeneracy pressure is 
strong enough to push against the infall of matter. Once nuclear densities are attained during this collapse, 
the matter rebounds outward in a shockwave. If this shockwave is successfully ejected, a neutron star is left behind; otherwise, the matter collapses back inward to form a black hole. 

Despite the ubiquity of CCSNe and their important role in the evolution of the universe, some details are difficult to study and also challenging to simulate fully in three dimensions. Given the low rate of a couple local CCSNe per century, this additionally complicates direct study through a single explosion. However, each CCSN produces $\sim$10$^{58}$ neutrinos of all flavors. There must exist, then, an ambient flux of these stellar messengers. This is the Diffuse Supernova Neutrino Background (DSNB), and we could be on the cusp of detecting it for the very first time.

\begin{figure}
\centering
\begin{minipage}{0.6\linewidth}
\centerline{\includegraphics[width=\linewidth]{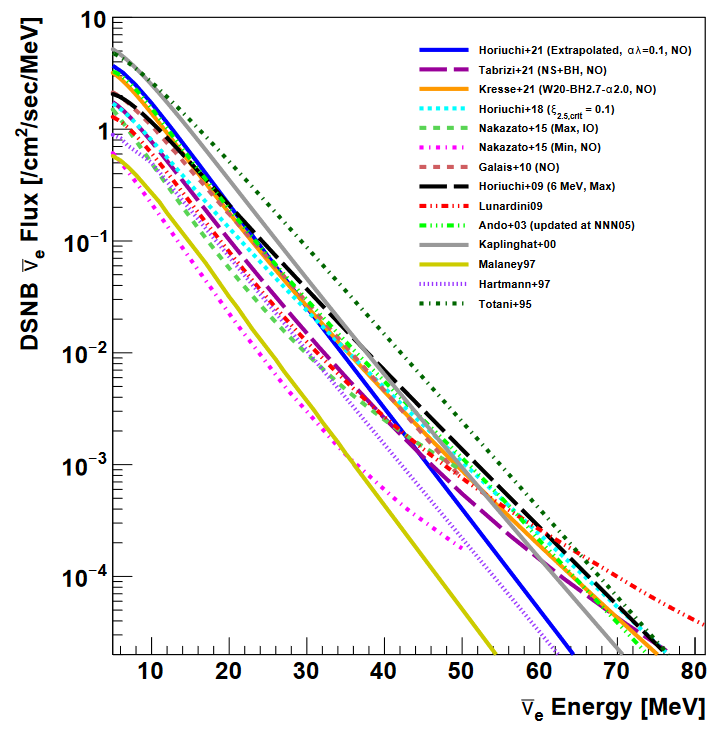}}
\end{minipage}
\caption[]{Selection of theoretical models for the $\bar{\nu}_e$ component of the DSNB flux.~\cite{dsnb21}}
\label{fig:dsnb_models}
\end{figure}

\section{Modeling Considerations}

There are many ways to model the DSNB, and Figure~\ref{fig:dsnb_models} demonstrates the wide variety of predictions possible. To illustrate what basic ingredients are necessary, we can consider a three-component model with

\begin{equation}
\frac{d\Phi_{\nu_\alpha}}{dE} = \int_{\rm CCSN} \int_{0}^{z_{\rm max}} R_{\rm CCSN}(z,M) \left[\frac{dF_{\nu_\alpha}(E(1+z), M)}{dM} \right] \left | c \frac{dt}{dz} \right| dzdM,
\label{eq:dsnb}
\end{equation}

\noindent where we integrate over the masses $M$ of CCSN-producing stars and redshift $z$. We make use of the rate of CCSNe ($R_{\rm CCSN}$), the emitted neutrino spectrum of flavor $\alpha$ ($F_{\nu_\alpha}$), and the cosmological evolution of the universe ($dt/dz$). With this, we understand that a DSNB measurement is not only a new source of neutrinos but also a new window into aspects of particle physics, astrophysics, and cosmology.

\begin{figure}
\begin{minipage}{0.44\linewidth}
\centerline{\includegraphics[width=\linewidth]{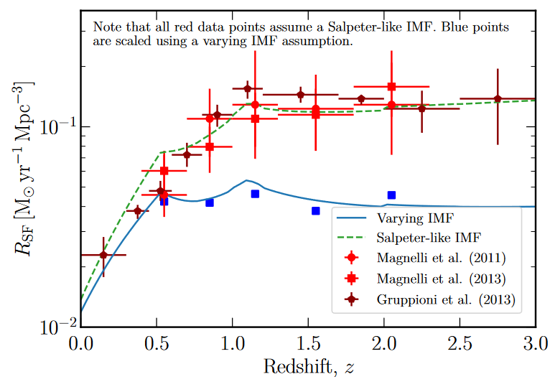}}
\caption[]{Star formation rate with various IMFs.~\cite{imfsfr22} \label{subfig:sfr}}
\end{minipage}
\hfill
\begin{minipage}{0.55\linewidth}
\centerline{\includegraphics[width=\linewidth]{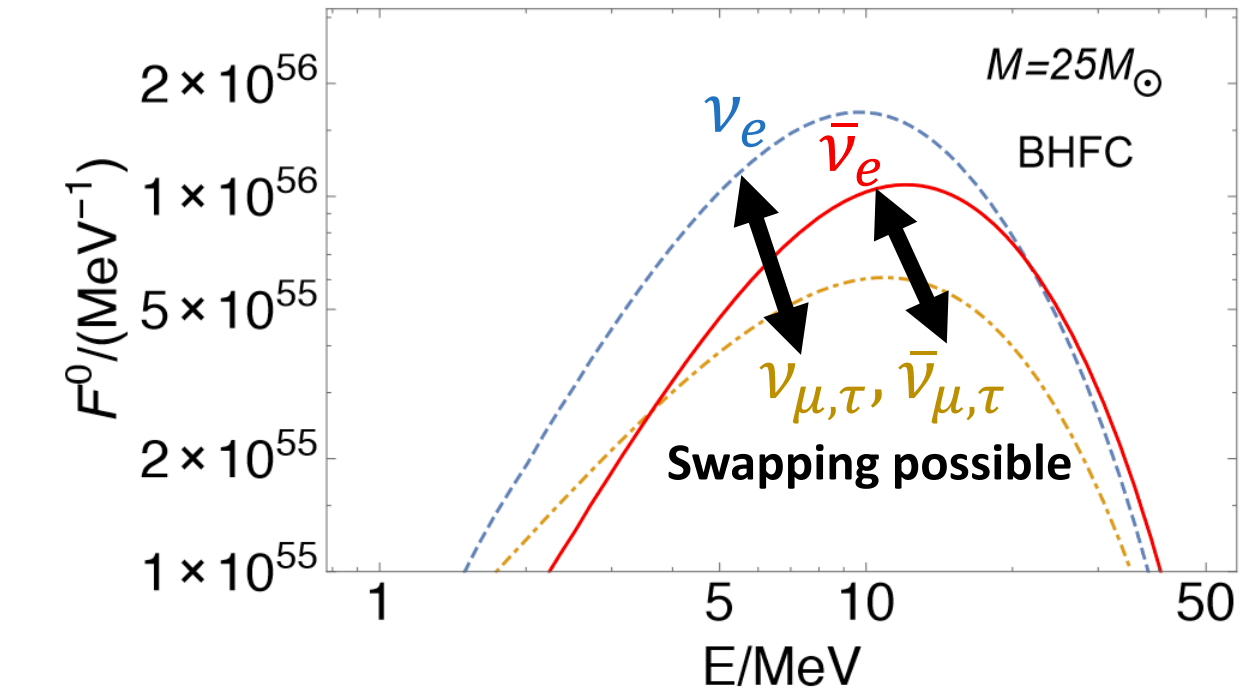}}
\caption[]{Neutrino emission spectra for various flavors from a black hole-forming CCSN.~\cite{priya17} Adapted to show spectral swapping. \label{subfig:swapping}}
\end{minipage}
\label{fig:sfr_swapping}
\end{figure}

It is conventional to define the rate of CCSNe as directly proportional to the stellar formation history of the universe. From Figure~\ref{subfig:sfr}, we note two important points about this rate $R_{\rm SF}$. The first is that the measurements often come with appreciable uncertainty, which leaves a large range of potential normalizations on $R_{\rm SF}$---one of the dominant uncertainties on the overall DSNB flux prediction. The second is that the assumption of how stellar masses are distributed in star-forming regions (the so-called ``initial mass function" or IMF) changes the interpretation of $R_{\rm SF}$ measurements.

\begin{figure}
\begin{minipage}{0.46\linewidth}
\centerline{\includegraphics[width=\linewidth]{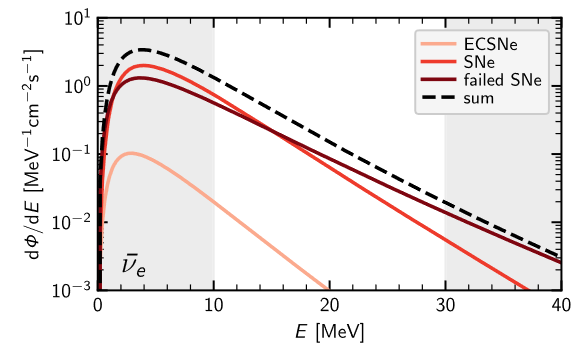}}
\caption[]{Contributions from various CCSN types to overall DSNB flux.~\cite{kresse21} \label{subfig:fBH}}
\end{minipage}
\hfill
\begin{minipage}{0.45\linewidth}
\centerline{\includegraphics[width=\linewidth]{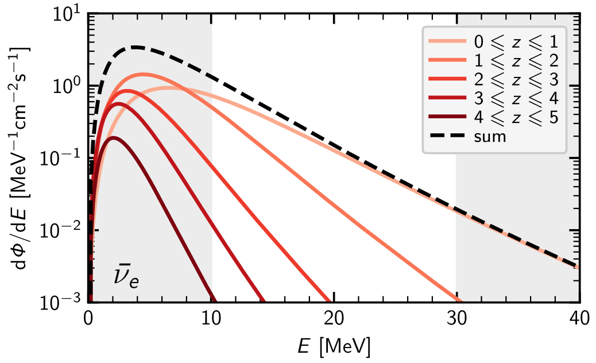}}
\caption[]{Contributions of CCSNe at various redshifts to overall DSNB flux~\cite{kresse21} \label{subfig:redshift}}
\end{minipage}
\label{fig:fBH_redshift}
\end{figure}

Next, we move to the neutrino emission flux. Due to the modified propagation of neutrinos through ultra-dense matter arising from the Mikheyev-Smirnov-Wolfenstein effect, the final emission spectra can swap flavors as seen in Figure~\ref{subfig:swapping}. This depends on the exact density profile of matter in CCSNe as well as the ordering of neutrino mass states. We often target $\bar{\nu}_e$ interactions in experiments, whose spectrum can swap with initially produced $\bar{\nu}_\mu$ and $\bar{\nu}_\tau$. Notably, this can lead to a decrease in total $\bar{\nu}_e$ flux. Furthermore, the fraction of CCSNe producing black holes also impacts the shape of this flux. In Figure~\ref{subfig:fBH}, we notice that the black hole-forming CCSNe (``failed SNe") generate hotter neutrino spectra compared to neutron star-forming CCSNe. This fraction is not well known observationally and could be constrained by a precise DSNB measurement in the future.

Finally, the cosmological expansion of the universe impacts the spectral contributions of CCSNe at varying tranches of redshift (see Figure~\ref{subfig:redshift}). Not only are the contributions diminished from the inverse squared distance, but the redshift of neutrino energies also moves the DSNB contributions to increasingly colder temperatures. As many experiments focus on DSNB neutrinos above around 10~MeV, the contributions beyond $z=2$ are often ignored in theoretical calculations. While the DSNB prediction is sensitive, in theory, to differences in cosmological models, the standard $\Lambda$CDM model with varying $H_0$ observations will not readily be probed in light of the other uncertainties.

In all, we see that there is a rich landscape of astrophysics, particle physics, and cosmology in a standard approach to modeling the DSNB. A first observation of its flux will be most sensitive to its normalization, while later measurements will narrow down finer structures in its spectrum. Fortunately, the experimental hunt is already underway.

\section{Experimental Searches}

A handful of experiments have contributed to the DSNB search in the past couple decades, and the next few years will welcome in an era of increasingly sensitive experiments. The Super-Kamiokande collaboration, in particular, has led the way with a new set of results in 2024.

\subsection{Super-Kamiokande}


The Super-Kamiokande (Super-K/SK) experiment is located in the Kamioka mine in Japan, comprising a 50-kton cylindrical water Cherenkov detector. To shield itself from cosmic ray muons, it has an overburden of 1000~m of rock (2700~m.w.e). Additionally, it makes use of 18~kton of its volume as an outer detector muon veto (lined with nearly 2000 8-in.\ PMTs), which encompasses the remaining 32~kton in its inner detector (lined with more than 11000 20-in.\ PMTs). In the 2020s, Super-K has performed two campaigns of gadolinium (Gd) loading to enhance neutron capture identification with the first happening in 2020~\cite{gd22} and the second in 2022~\cite{gd24} to achieve 0.01\% and 0.03\% Gd-loading, respectively.

Super-K searches for inverse beta decay (IBD) signatures to detect the DSNB,

\begin{equation}
    \bar{\nu}_e + p \to e^+ + n,
\end{equation}

\noindent for which the positron generates an immediate ``prompt" Cherenkov signal and the neutron is subsequently captured after thermalization to create a ``delayed" signal. Therefore, not only is effective prompt signal identification important but also neutron identification. Without Gd-loading, neutron captures in pure water emit one 2.2~MeV gamma with characteristic timescale 200~$\mu$s. With 0.01\% (0.03\%) Gd-loading, 50\% (75\%) of true captures happen on Gd nuclei with characteristic timescales of $\sim$120~$\mu$s ($\sim$60~$\mu$s) and multiple gamma emission with $\sim$8~MeV total. A pure-water DSNB search~\cite{dsnb21} was performed with two parallel approaches: an unbinned energy spectrum fit and an energy-binned analysis. In its Gd-era, Super-K performed a first search~\cite{dsnb23} with an energy-binned analysis. In the summer of 2024, the newest Super-K results with both Gd-loading periods were released~\cite{harada24,beauchene24,rogly24,santos24}.

One major update was the use of two machine learning neutron identification tools, one a Boosted Decision Tree (BDT) and the other an artificial Neural Network (NN). Another important update was the introduction of a new background reduction step targeting especially atmospheric neutral current quasi-elastic (NCQE) backgrounds, usually dominant below around 20~MeV. Using an observable traditionally characterizing the ``multiple scattering goodness" (MSG) of solar neutrinos, the NCQE events were reduced by up to an order of magnitude further with a corresponding loss of around 15\% in DSNB signal across the energy region of interest.~\cite{moriond24} This is because the MSG variable is effective at probing the sub-structure of Cherenkov cone patterns on the PMTs and can therefore separate single-cone IBD events from overlapping cone events of NCQE coming from multi-gamma emission of the knockout nucleons.~\cite{ncqe24}

\begin{figure}
\begin{minipage}{0.42\linewidth}
\centerline{\includegraphics[width=\linewidth]{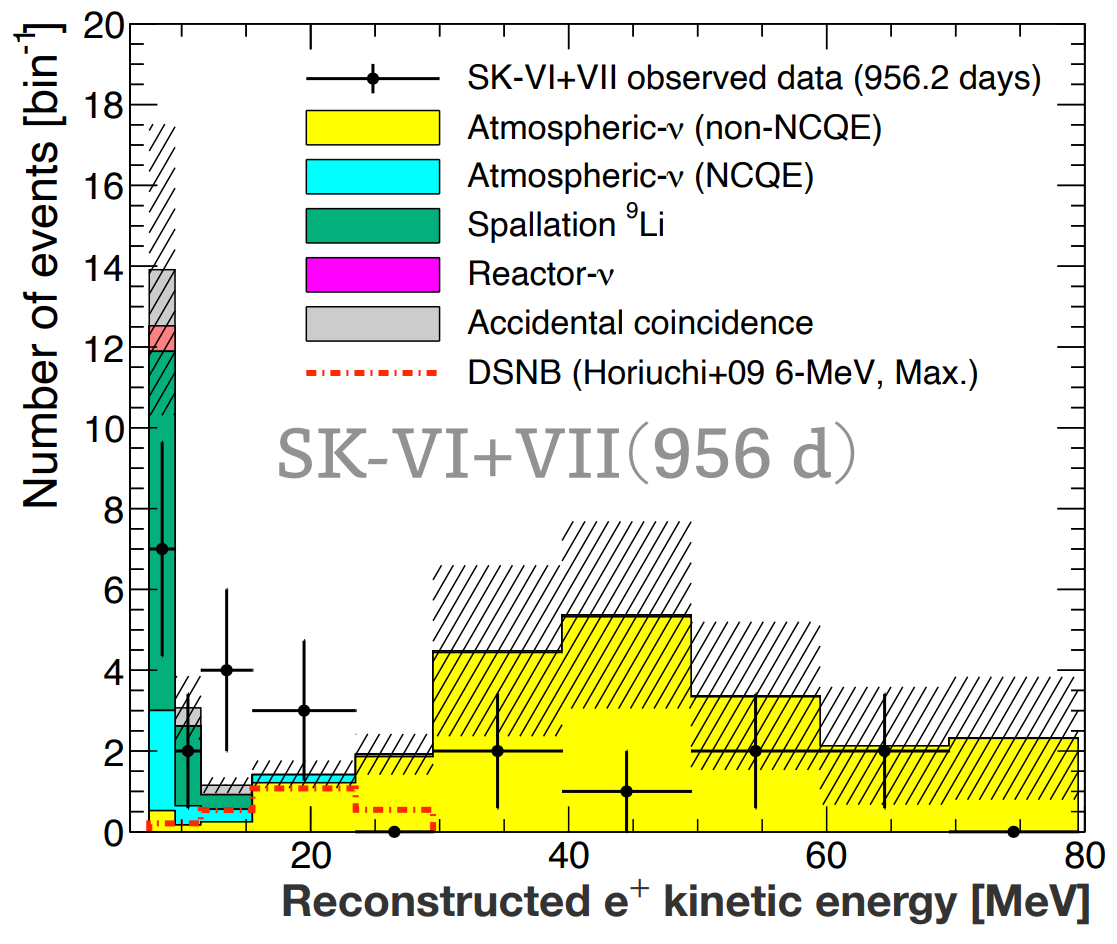}}
\caption[]{Combined energy-binned spectrum for SK-VI/VII periods.~\cite{harada24} \label{subfig:binned_results}}
\end{minipage}
\hfill
\begin{minipage}{0.54\linewidth}
\centerline{\includegraphics[width=\linewidth]{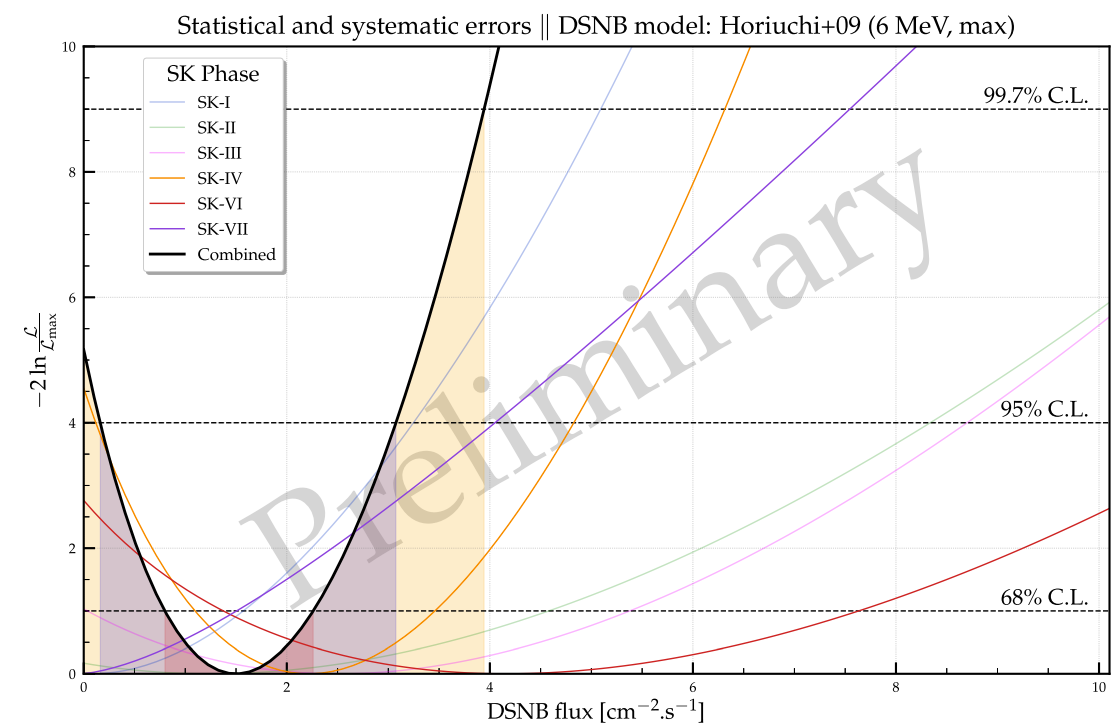}}
\caption[]{Spectral fitting result~\cite{harada24,beauchene24,rogly24} for one DSNB theoretical model, showing all individual SK phases considered and their combination (black, solid). \label{subfig:spectral_results}}
\end{minipage}
\label{fig:dsnb_results}
\end{figure}

Both an unbinned energy spectrum fit and an energy-binned analysis were carried out. A light excess over background-only predictions was observed in the binned analysis for the Gd-era, depicted in Figure~\ref{subfig:binned_results}. In the spectrum fit, approximately a $2.3\sigma$ rejection of a background-only hypothesis was found across all models studied when combining all SK phases. An example DSNB model's profile likelihood ratio is shown in Figure~\ref{subfig:spectral_results}.

\subsection{Jiangmen Underground Neutrino Observatory}

The Jiangmen Underground Neutrino Observatory (JUNO) is in its final stages of construction in Guangdong province in China. It will be a 20~kton liquid scintillator detector with an overburden of 700~m of rock (1800~m.w.e.). As part of its cosmic ray muon veto approach, its central inner detector will be held within a cylindrical water tank containing 2400 20-in.\ PMTs. It will aim for highly effective energy reconstruction, making use of almost 18000 20-in.\ PMTs with another 25600 3-in.\ PMTs filling the gaps in-between. The collaboration currently anticipates data-taking to begin in 2025.

In recent years, JUNO has detailed a plan~\cite{juno22} for its DSNB search. Its energy search window will be similar to Super-K, constrained below by spallation and reactor $\bar{\nu}_e$ and above by charged-current atmospheric neutrino interactions. The collaboration anticipates highly effective cosmic ray muon tagging and spallation reduction. Also, it expects to easily remove ``fast neutrons" entering the detector from untagged muons interacting in the surrounding rock. Finally, the impact of atmospheric neutral current interactions should be reduced by studying the PMT pulse shape from multi-component signals of the atmospheric neutrino scattering off ${}^{12}$C.

\subsection{Hyper-Kamiokande}

The Hyper-Kamiokande (Hyper-K/HK) experiment is under construction with plans to start taking data in 2027. It will continue the legacy of the Kamiokande series with around 8 times the fiducial volume of Super-K. Initially, Hyper-K will not be Gd-loaded. One crucial difference between Super-K and Hyper-K is the smaller overburden of Hyper-K: At a depth of 650~m under rock (1750~m.w.e.), the cosmic ray muon flux will be a bigger hurdle for the DSNB search than in Super-K. A preliminary strategy has been proposed~\cite{hk18} to estimate this flux and reduce its impact on low-energy physics analyses. Ongoing efforts aim to update sensitivity studies ahead of the start of the experiment.

\subsection{LUX-ZEPLIN}

The LUX-ZEPLIN (LZ) experiment located at the Sanford Underground Research Facility in the United States is a 7~ton liquid xenon detector. While its primary purpose is to detect dark matter directly, it is also sensitive to coherent elastic neutrino-nucleus scattering (CE$\nu$NS). While these kinds of limits are generally weaker than $\bar{\nu}_e$ limits from IBD searches, LZ set a 90\% C.L.\ upper limit on the DSNB flux of muon and tau flavors at 686--826~cm$^{-2}$s$^{-1}$ for energies above 19.3~MeV.~\cite{lz24} These are of the same order of magnitude of Super-K limits for the same flavors.

\subsection{Experimental Sensitivity Outlook}

\begin{figure}
\centering
\begin{minipage}{0.8\linewidth}
\centerline{\includegraphics[width=\linewidth]{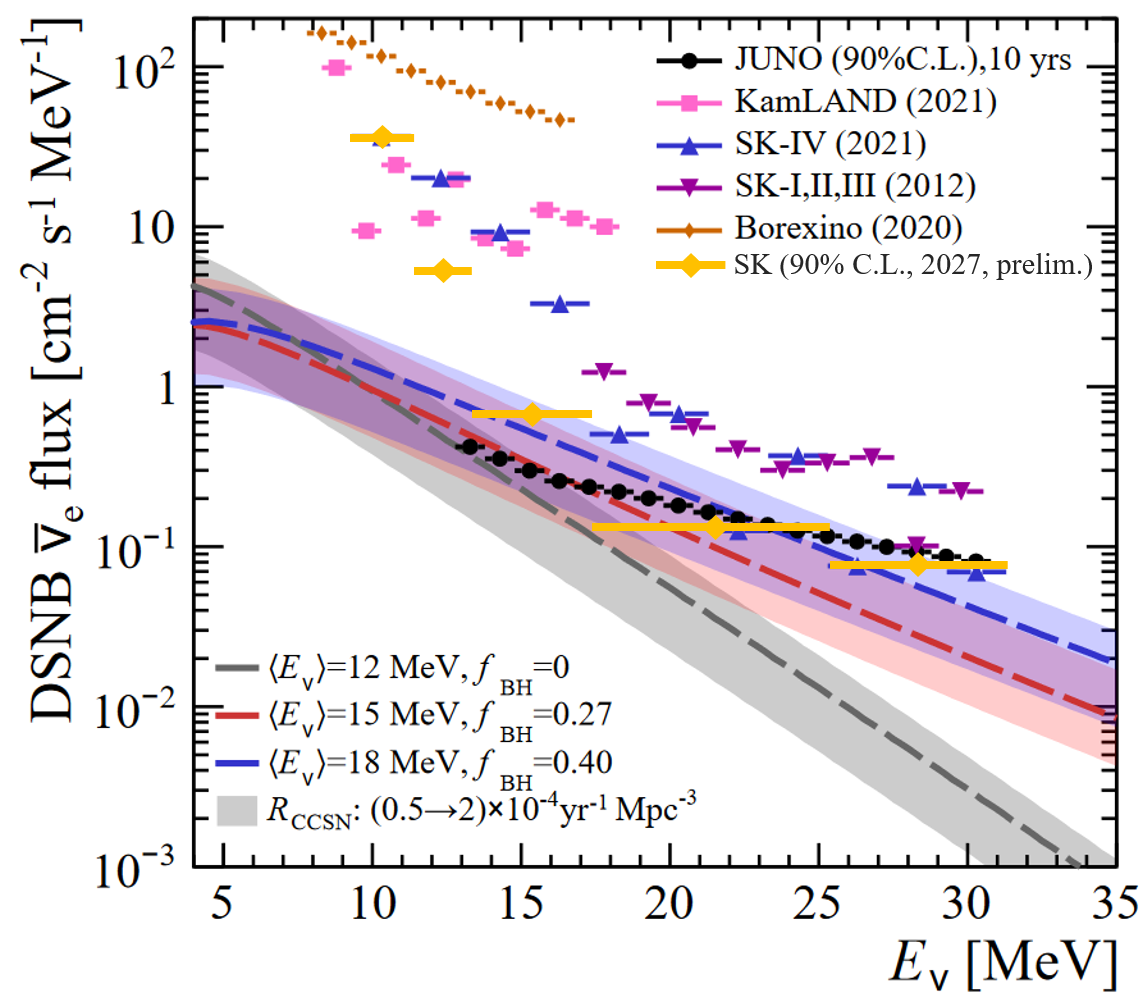}}
\end{minipage}
\caption[]{DSNB upper limit comparison between JUNO (expected), KamLAND (observed), SK-IV (observed), SK-I/II/III (observed), Borexino (observed), and SK-IV/Gd~\cite{santos24} (expected). Adapted from JUNO.~\cite{juno22}}
\label{fig:upper_limit_compare}
\end{figure}

In Figure~\ref{fig:upper_limit_compare}, we show both observed and expected upper limits on the DSNB $\bar{\nu}_e$ flux from several experiments. Notably, the Super-K and JUNO experiments will be running at the same time, and Hyper-K will join soon after. We do not include any estimate for Hyper-K sensitivity as this is still an ongoing study. For Super-K, we understand that the sensitivity at higher energies will be achieved faster than the projected JUNO sensitivity. However, at lower energies, JUNO is expected to perform well due to significant spallation background reduction. In the long term, the Super-K/Hyper-K and JUNO experiments could be complementary across the full energy search region.

\section{New Physics with the DSNB}

While we have already outlined a standard set of uncertainties on the theoretical modeling of the DSNB spectrum, it is equally as feasible to imagine new physics scenarios that can be uniquely studied with the DSNB.

\begin{figure}
\begin{minipage}{0.44\linewidth}
\centerline{\includegraphics[width=\linewidth]{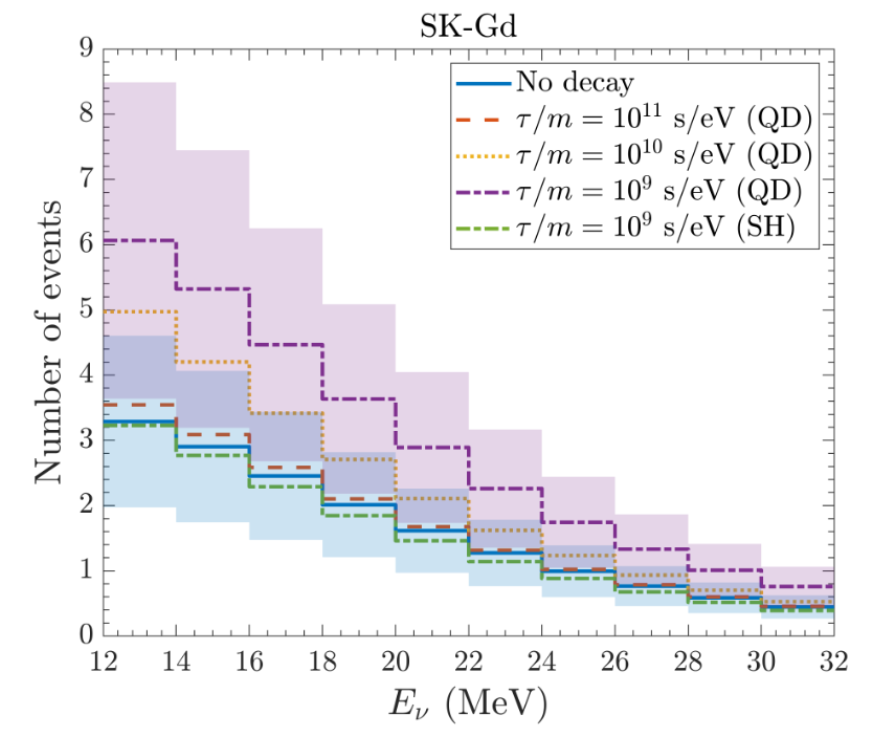}}
\caption[]{DSNB events in SK-Gd as a function of neutrino non-radiative decay scenarios.~\cite{pib23} \label{subfig:nu_decay}}
\end{minipage}
\hfill
\begin{minipage}{0.48\linewidth}
\centerline{\includegraphics[width=\linewidth]{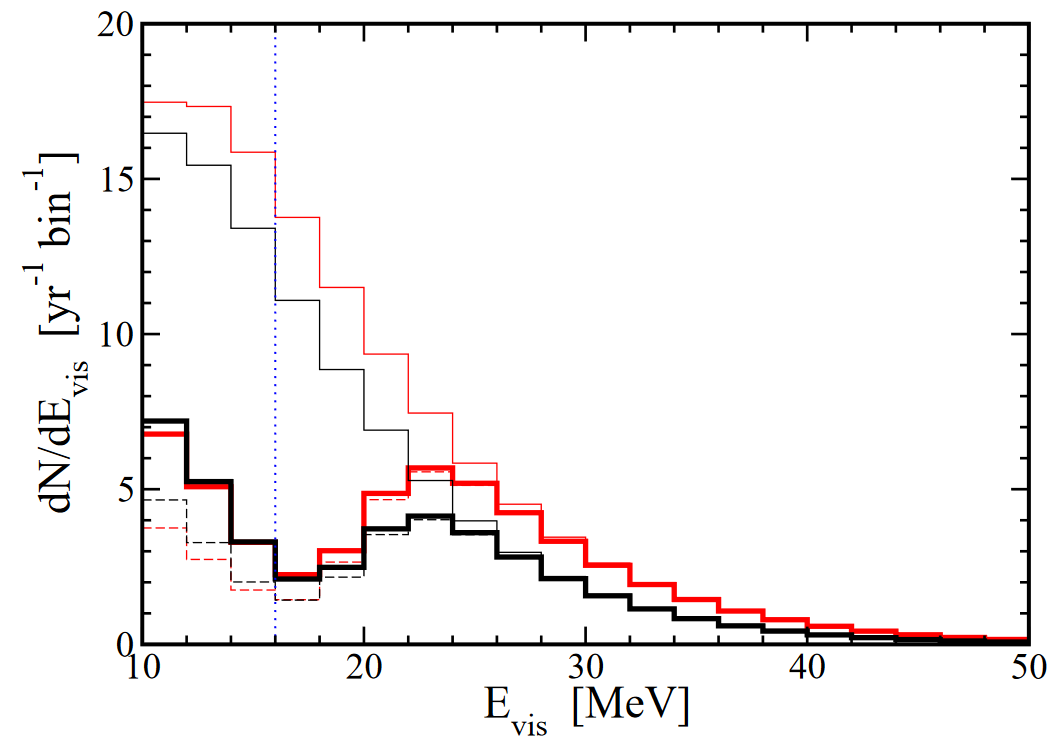}}
\caption[]{DSNB event rate in HK with (thick) and without (thin) strong $\nu_\tau$-DM coupling in normal (black) and inverted (red) mass hierarchy.~\cite{farzan14} \label{subfig:dm_resonance}}
\end{minipage}
\label{fig:new_physics}
\end{figure}

\subsection{Neutrino Non-radiative Decay}

In a scenario for which neutrinos can undergo non-radiative decay from heavier to lighter mass states along with a (nearly) massless scalar, the nominal DSNB flux prediction can be significantly modified.~\cite{pib23} For example, when the heaviest and lightest mass states $m_h$ and $m_l$ satisfy $m_h-m_l \gg m_l \simeq 0$, the decays of ``heavier" flavor states in the normal mass ordering go to the ``lighter" electron flavor, enhancing the DSNB $\bar{\nu}_e$ signal as seen in Figure~\ref{subfig:nu_decay}. There is, instead, a $\bar{\nu}_e$ suppression in inverted ordering.

\subsection{Dark Matter Resonances}

Since DSNB neutrinos detected on Earth have traveled significant distances, there are ample opportunities for them to interact in novel ways with the surrounding universe. One such case is through a resonance arising from a general coupling of neutrinos to dark matter. Due to constraints from meson decay experiments on this kind of coupling from electron and muon flavors, one can focus on a demonstrative 20~MeV resonance in which DSNB neutrinos are absorbed from strong tau flavor coupling.~\cite{farzan14} Depending on the exact realization of this resonance, a dip in the $\bar{\nu}_e$ spectrum could be observed in an experiment like Hyper-K (see Figure~\ref{subfig:dm_resonance}).

\begin{figure}
\centering
\begin{minipage}{0.8\linewidth}
\centerline{\includegraphics[width=\linewidth]{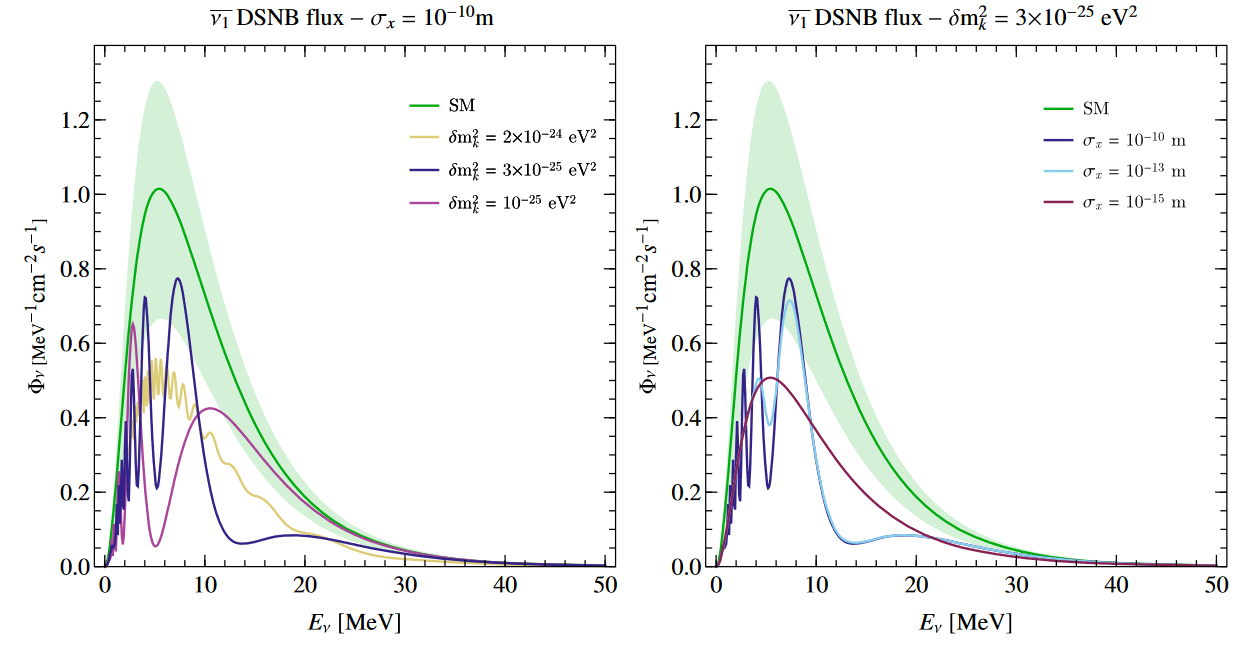}}
\end{minipage}
\caption[]{DSNB flux predictions with new ``exotic" physics.~\cite{degouvea20} (Left) Pseudo-Dirac neutrinos with various mass-squared differences from adding sterile neutrinos. (Right) Various wave packet sizes in the presence of sterile neutrino mixing.}
\label{fig:exotic}
\end{figure}

\subsection{Sterile Neutrinos and Wave Packet Considerations}

For pseudo-Dirac neutrinos with sterile states, there is a possibility for non-trivial active-sterile neutrino mixtures in the presence of significantly small mass-squared differences or even finite wave packet sizes.~\cite{degouvea20} This kind of new physics tends to deplete the predicted DSNB flux and introduce new ``wiggles" in the spectrum, shown in Figure~\ref{fig:exotic}. Once again, given the large distances traversed by DSNB neutrinos before detection, such scenarios in which large oscillation lengths arise from small mass-squared differences are capable of being probed.

\section{Conclusion}

The DSNB is an exciting new source of astrophysical neutrinos that we have yet to detect. It will be capable of probing topics ranging from astrophysics, particle physics, cosmology, and even beyond Standard Model frameworks. A suite of experiments have searched for this signal, and we are entering the most sensitive era to date between Super-K, Hyper-K, and JUNO. Soon, we can hope to go from the years-long \textit{search} for the DSNB to DSNB-\textit{driven} physics.




\section*{References}

\begin{thebibliography}{99}

\bibitem{dsnb21} K. Abe {\it et al}, \Journal{\PRD}{104}{122002}{2021}

\bibitem{imfsfr22} J.J. Ziegler {\it et al}, \Journal{\it MNRAS}{517}{2471}{2022}

\bibitem{priya17} A. Priya and C. Lunardini, \Journal{\it JCAP}{11}{031}{2017}

\bibitem{kresse21} D. Kresse {\it et al}, \Journal{\APJ}{909}{169}{2021}

\bibitem{gd22} K. Abe {\it et al}, \Journal{\NIMA}{1027}{166248}{2022}
\bibitem{gd24} K. Abe {\it et al}, arXiv:2403.07796

\bibitem{harada24} M. Harada, {\it Proceedings of Neutrino 2024} (2024)

\bibitem{beauchene24} A. Beauch\^ene, {\it Proceedings of Neutrino 2024} (2024)
\bibitem{rogly24} R. Rogly, {\it Proceedings of Neutrino 2024} (2024)

\bibitem{dsnb23} M. Harada {\it et al}, \Journal{\PRL}{951}{L27}{2023}

\bibitem{santos24} A.D. Santos {\it et al}, {\it Proceedings of Neutrino 2024} (2024)

\bibitem{moriond24} A.D. Santos, arXiv:2405.07900

\bibitem{ncqe24} S. Sakai {\it et al}, \Journal{\PRD}{109}{L011101}{2024}

\bibitem{juno22} A. Abusleme, \Journal{\it JCAP}{10}{033}{2022}

\bibitem{hk18} K. Abe {\it et al}, arXiv:1805.04163

\bibitem{lz24} Q. Xia, arXiv:2412.15886

\bibitem{pib23} P. Iva\~nez-Ballesteros and M.C. Volpe, \Journal{\PRD}{107}{023017}{2023}

\bibitem{farzan14} Y. Farzan and S. Palomares-Ruiz, \Journal{\it JCAP}{06}{014}{2014}

\bibitem{degouvea20} A. de Gouv\^ea {\it et al}, \Journal{\PRD}{102}{123012}{2020}

\end{thebibliography}
\end{document}